\newcommand{\be}{\begin{equation}}
\newcommand{\ee}{\end{equation}}

\documentclass[aps,prd,twocolumn,showpacs]{revtex4}
\usepackage{amssymb}
\usepackage{graphicx}
\usepackage{dcolumn}
\usepackage{bm}
\usepackage{amsmath}
\usepackage{color}
\begin{document}

\title{Can Quantum Mechanics explain the Evolution of the Universe?}

\author{Vladlen G. Shvedov}

\address{Laser Physics Centre, Research School of Physics and Engineering, The Australian National University, Canberra ACT 0200, Australia}

\begin{abstract}
This manuscript deals with a model of the evolution of an event space represented by the fundamental solution of a N-dimensional generalized Schr\"odinger equation for free matter. Specifically this solution can be applied to describe the 3D space evolution of the Universe in the forward direction in time. The model which is based on the presented solution is close to the well known  Inflation theory, but  is  nonsingular, does not violate the conservation laws and is finite throughout the whole space at any moment of time. According to this model, the evolution progresses non-uniformly with a positive acceleration. Moreover, the model correctly approaches Hubble's law in the long-time limit. It is also shown that any source functions which describe substance and its physical fields set space topology of the Universe, but do not influence the general dynamics of its evolution.
\end{abstract}

\pacs{42.25.Bs, 98.80.Qc, 98.80.Bp}

\maketitle

Research on  electromagnetic waves (EMW) has had  an enormous impact on the development of modern physics.
Two fundamental concepts  of modern physics: the special theory of relativity and quantum mechanics (QM), must conform  to principles of a constancy of velocity  and corpuscularity   of light. The deep analogies between wave optics and QM are well known ~\cite{Fermi}. Nevertheless, the  rapid development of QM in the last hundred years has shifted  classical representation of  EMW to  the second plan. In fact, the approximate solutions of Maxwell's equations describing coherent waves in the directed beams were found only in second part of the last century ~\cite{Kog}. They had appeared from the necessity of the description of a set of the experimental facts connected with laser radiation. However, a new puzzle had emerged. For the correct description of a scalar monochromatic (i.e. homogeneously polarized ~\cite{bible}) wave $\psi$, with finite energy and preferential direction of propagation it is necessary to abandon the spatially invariant Helmholtz wave equation  ~\cite{bible}: $({\bf\nabla}^2+k^2)\psi=0$. Here  ${\bf\nabla}=\partial_{x}{\bf e}_{x}+\partial_{y}{\bf e}_{y}+\partial_{z}{\bf e}_{z}$; $\partial_{u}\equiv \partial/\partial u$; ${\bf e}_{x}$, ${\bf e}_{y}$, ${\bf e}_{z}$ are the Cartesian unit vectors, $k=\omega/c$ is the wave number, $\omega$ is the  frequency of light and $c$ is the speed of light. The simplest approximation to the Helmholtz equation may be found by assuming  that wave field evolves  in one spatial direction (for example $z$) more slowly than in others ~\cite{Lax}: $\mid\partial_{z}^{2}{ \psi}\mid\ll\mid\partial_{z}{\psi}\mid$. It means that in this case the $z$ axis represents a preferential direction similar to the time arrow in mechanical processes. Then the wave equation with a good accuracy becomes: ~\cite{book,Sei}:
\begin{equation}
\label{2}
\nabla_{\bot}^{2}\psi=-2ik\partial_{z}\psi,
\end{equation}
where ${\bf\nabla_\bot}=\partial_{x}{\bf e}_{x}+\partial_{y}{\bf e}_{y}$. This equation  corresponds exactly to the Schr\"odinger equation of a reduced dimension 2+1 for the wave function $\psi$ of the free matter with equivalent mass $\tilde m=k\hbar/c$ ($\hbar$ is the Planck constant) if the spatial coordinate of propagation $z$ is replaced by time coordinate $ct$. The reduced dimension of Eq.(\ref{2}) relative to the full 3+1 dimensional space-time Schr\"odinger equation for free evolving matter of mass $m$:
\begin{equation}
\label{3}
\nabla^{2}{\mit\Psi}=-2im/\hbar\:\partial_{t}\mit\Psi,
\end{equation} leads to a different physical interpretation of the corresponding solutions of Eq.(\ref{2}) and Eq.(\ref{3}), but the structure of these solutions should be similar.
Therefore, before proceeding any further it will be instructive to discuss the well-known solutions of Eq.(\ref{2}) describing propagation of EMW.

{\em EMW solutions}. A prominent feature of all known finite solutions of the Eq.(\ref{2}) is that they posses Gaussian kernel ~\cite{book,Sei}:
\begin{equation}
\label{4}
G(\rho,z)= \frac{\sqrt{2}A}{\sqrt{\pi}w_0s(z)}\exp{\left(-\frac{\rho^2}{w_{0}^2s(z)}\right)},
\end{equation}
where $s(z)=1+2iz/\left(kw_{0}^2\right)$, $\rho=\sqrt{x^2+y^2}$ is the radial coordinate of an arbitrary point, $A$ is a constant field amplitude and $w_0$ is a localization parameter of the wave in plane $z=0$. The function (\ref{4}) is the fundamental solution of the Eq.(\ref{2}) and plays a key role in all its physical solutions. It describes own evolution of the wave function $\psi$ and does not depend on source functions in general solutions of the Eq.(\ref{2}).
\begin{figure}
\centerline{\includegraphics[width=1\columnwidth]{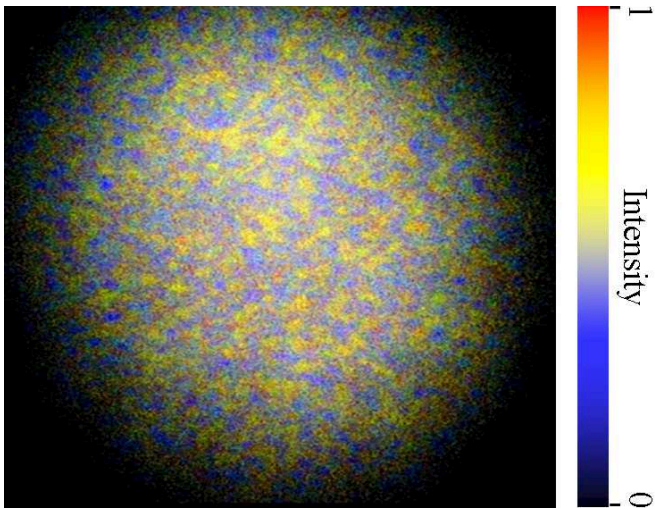}}
\caption{\label{fig1} Experimental photo of the intensity distribution inside a coherent $\lambda=633 ~ \rm {nm}$ Gaussian light field (\ref{2}) modulated with a far-field speckle pattern. The topological charges (the points with zero intensity) are distributed uniformly on average, but they can form local clusters which are introduced by the large-scale structure of the intensity.}
\end{figure}
Indeed, general finite solution of Eq.(\ref{2}) can be represented as a linear combination of the following expressions ~\cite{book,Kis}
\begin{equation}
\label{5}
\psi(x,y,z)=f(x,y,z)G(x,y,z),
\end{equation}
where $f(x,y,z)$  is the modulating function which is not satisfying  the Eq.(\ref{2})in contrast to background function $G(x,y,z)$ ~\cite{Kis}. By hanging  variables in Eq.(\ref{5}) as $\zeta=s(z)w_0$, $X=x/\zeta$, $Y=y/\zeta$ one can show that the modulating function satisfies  the following differential equation: $(\partial^2_{X}+\partial^2_{Y}-(1/2)ik\zeta^2\partial_{\zeta})f(X,Y,\zeta)=0$ which admits the following separation of variables: $f(X,Y,\zeta)=F(X,Y)Z(\zeta)$, where $Z(\zeta)=\exp (-K^2/k^2w^2_0s(z))$ and $K$ is a complex constant ~\cite{Kis}. Now the  function $F(X,Y)$ satisfies a two-dimensional Helmholtz wave equation:
\begin{equation}
\label{1}
(\partial^2_{X}+\partial^2_{Y}+K^2)F=0;
\end{equation}
In many cases ~\cite{book,Sei} the constant $K$ is equal to zero and the Eq.(\ref{1}) leads to Laplace's equation for the 2D field of a source: $(\partial^2_{X}+\partial^2_{Y})F=0$. In that case the function $f(X,Y)$ does not depend on $\zeta$ and $f(X,Y)=F(X,Y)$. When the field $F$ has a cylindrically symmetric distribution its Laplace's equation can be present in the form:
\begin{equation}
\label{6}
\nabla_{\bot}^{2}F(\rm p,\varphi)=0.
\end{equation}
where $\rm p=\mit\rho/\zeta$; $\rho$, $\varphi$, $z$ are cylindrical coordinates.
Known  solutions of Eq.(\ref{6}) represent discrete sets of  polynomial functions~\cite{Sei}. These  functions determine the topology of the background wave (\ref{6}) without  influencing the general character of its evolution. This can be illustrated by an example of the elementary solution of Eq.(\ref{7}):
\begin{equation}
\label{7}
F(\rm p,\varphi)=\rm p^{\mit|l|}\exp(\mit il\varphi),
\end{equation}
where $l=0;\pm1;\pm2;..$. The discrete character of Eq.(\ref{7}) forbids a smooth transition between groups of solutions with different $l$ number. Such a transition can only be realized as discrete leaps between different states. Along the line $\rho=0$ the phase of function (\ref{7}) is  singular while its modulus is identically zero when $l\ne 0$. Note that Eq.(\ref{7}) has no physical meaning without the background function (\ref{4}) since it leads to the intensity divergence if  $\rho\to\infty$. Only fundamental Gaussian function (\ref{4}) is responsible for an energy finiteness  of  directed spatially-unlimited waves described by Eq.(\ref{2}) ~\cite{book,Sei}. Nevertheless, the source function (\ref{7}) gives perfect information about the topological structure of the solution (\ref{5}): \begin{equation}
\label{8}
\psi_l=\frac{\sqrt{2^{|l|}}}{\sqrt{|l|!}}\frac{\rho^{|l|}}{\zeta^{|l|}}\exp(il\varphi)G(\rho,z).
\end{equation}
The factor $\sqrt{2^{|l|}}/(\sqrt{|l|!}\zeta^{|l|})$ appears from the requirement of a continuity of the energy flux in Eq.(\ref{8}) in the $z$-direction: $\iint\limits_{S_{\bot}\to\infty}\psi_l\psi_l^*\,d S_{\bot}=1$. Here $S_{\bot}$  is the area in  a plane orthogonal to the wave propagation direction, and the star denotes  complex conjugation. The index $l$ is the topological charge defining the topological structure of the wave surface ~\cite{Nye74}. It is characterized by a singularity line which extends through the wave field (\ref{8}). As the topological part of the Eq.(\ref{8}) contains only in function (\ref{7}) the  topological charges do not rigidly spatially connected with the  background function (\ref{4}).
The line of singularity in Eq.(\ref{7}) coincides with the line of amplitude maximum of Eq.(\ref{4}) only in specific case of Eq. (\ref{8}). Generally, a wave field may contain  a large number  of topological charges. Their  trajectories can be complex and even entangled as, for instance, in speckle fields (see FIG.1) ~\cite{Den, My}. Moreover, feasible modulating functions of sources can not to be scalar functions. In special cases they can satisfy only vectorial Laplace's equations $\nabla_{\bot}^2{\bf F}({\rm p},\varphi)=0$. Two types of such  singular modulating functions are known in wave optics: ${\bf F}_\rho={\rm p}{\bf e}_\rho$ and ${\bf F}_\varphi={\rm p}{\bf e}_\varphi$ representing radially and azimuthally polarized field.

Irrespective of their specifics   modulating functions can substantially change the topological structure of the wave field (\ref{7}) and, as a consequence, the intensity distribution(\ref{7}). However, it should be stressed that  they can physically exist only together with  the background function ((\ref{4}) and do not affect  the general character of the  evolution of the latter. Hence, one can speak of the function (\ref{4}) generating  a "space" for existence of topological objects in a wave field.
{\em QM solutions.} Analysis of the solutions of (2+1)D Schr\"odinger Eq.(\ref{2}) leads  to one important conclusion. If the fundamental solution (\ref{4}) of Eq.(\ref{2}) describes the evolution of 2D "space" of a wave function with finite energy in the third propagation dimension, thus, there should exists  an analogous  solution of the (n+1)D generalized Schr\"odinger equation:
\begin{equation}
\label{21}
\sum_{i=1}^n \frac{\partial^2}{\partial x_i^2}{\mit\Psi}=-2ic\frac{m}{\hbar}\frac{\partial}{\partial x_{n+1}}\mit\Psi,
\end{equation}
where $x_{n+1}$ denotes the coordinate of the preferred direction of the wave propagation, c is a velocity of the wave propagation along $x_{n+1}$ direction.  The general solution of Eq.(\ref{21}) reads:
\begin{equation}
\label{22}
{\mit\Psi_0(r_n,x_{n+1})}={\left(\rho_0\xi\sqrt{\pi/2}\right)^{-n/2}}\exp{\left(-\frac{r_n^2}{\rho_{0}^2\,\xi}\right)},\end{equation}
where $r_n$ is the n-dimension  space coordinate $r_n^2=\sum_{i=1}^nx_i^2$, $\xi=1+2i\hbar x_{n+1}/(m\rho_0^2)$, $\rho_0$ is the initial compression or the radius of $n$-dimensional sphere where the amplitude of the  wave function equals  $e^{-1}$ of its peak  at the moment $x_{n+1}=0$.

In the case of (3+1)D space-time world Eq.(\ref{21}) is just  the well known Schr\"odinger equation (\ref{3}) with propagation coordinate $x_{n+1}=ct$ and its  solution (\ref{21}) becomes:
\begin{equation}
\label{9}
{\mit\Psi_0(r,t)}= {\left(\rho_0\xi(t)\sqrt{\pi/2}\right)^{-3/2}}\exp{\left(-\frac{r^2}{\rho_{0}^2\,\xi(t)}\right)},
\end{equation}
 where $r^2=x^2+y^2+z^2$, $\xi(t)=1+2i\hbar t/(m\rho_0^2)$, $m$ is mass of matter, $\rho_0$ is the initial compression at the moment $t=0$. Formally the Eq.(\ref{9}) resembles  the well known QM propagator of wave packet which describes probability amplitude of the position of a point particle in the point $x_2$ at the moment $t_2$ when known its position $x_1$ at the moment $t_1$: $G_{+}^{(0)}(x_2,t_2;x_1,t_1)=\left(m/(2\pi\hbar i(t_2-t_1)\right)^{3/2}\exp{\left(im(x_2-x_1)^2/2\hbar(t_2-t_1)\right)}$ ~\cite{qm}.  On the contrary, the wave function (\ref{9}) has no connection with any point particle and concept of that particle itself has no physical meaning in Eq.(\ref{9}). The function (\ref{9}) is defined in the full space-time domain: $r\ge0;\, -\infty<t<+\infty$ and is not  localized anywhere.

 The choice of the time reference point $t^\prime=t\pm t_0$, where $t_0$ is an arbitrary constant, does not influence the direction of the function evolution (\ref{9}). The unidirectional positive time arrow in the whole  space is defined by the sign of the time derivative in Eq.(\ref{3}). Therefore, the evolution of the function (\ref{9}) follows one time direction in any space point. Note that the solution (\ref{9}) is nonsingular and  is finite throughout the whole space at any moment of time. Formally this means  that the modulus square (intensity) of the wave function  integrated over the whole volume  of the space $V$ ($dV=dxdydz$), is conserved  in  time:
 \begin{equation}
\label{10}
\iiint_{V\to\infty}{\mit\Psi_0\mit\Psi_0^*}\,d V = 1.
\end{equation}
In the QM interpretation Eq.(\ref{10}) signifies the fact that probability of all events which are defined by the function (\ref{9}) equals  one  at any time. On the other hand,  Eq.(\ref{9}) describes spatial evolution in time of a free matter in the center-of-mass system. Therefore  expression (\ref{9}) is represents fundamental solution  in (3+1)D space-time, as well as the expression (\ref{4}) is the fundamental solution  in (2+1)D space. For the given parameters mass $m$ and initial compression $\rho_0$ this function is unique in the (3+1)D space-time. This fact suggests that  (\ref{9}) is actually the fundamental expression describing the evolution of the space of events of free matter in time. In our physical world such space is nothing else but the space of the  Universe. Now let us consider what conclusions can be drawn from such interpretation of the  basic evolutionary solution (\ref{9}). First of all, it follows from  Eq.(\ref{9}) that the space of events is lying on the wavefront surface $r^2mc/(2\hbar R(t))-(3/2)\arctan(t/\tau_c)=const$, were $\tau_c$ is a  constant defined as
\begin{equation}
\label{12}
\tau_c=m\rho_0^2/(2\hbar),
\end{equation}
and
\begin{equation}
\label{23}
R(t)=ct(1+m^2\rho^4_0/(4\hbar^2t^2)),
\end{equation}
is the radius of curvature of the wavefront.

The wave function (\ref{9}) is antisymmetric in time with respect to the origin of coordinates $t=0$, $r=0$. It monotonously contracts in a time interval  $-\infty<t\le0$ and monotonously expands in an interval $0\le t<+\infty$ according to  the hyperbolic law
\begin{equation}
\label{11}
r=r_0\sqrt{1+t^2/\tau_c^2},
\end{equation}
where $r_0$ is the coordinate of an arbitrary fixed point on the wavefront surface at the moment $t=0$, $r$ is the coordinate of the same point at the moment $\pm t$.
The function $r(r_0, t)$ is an infinitely differentiable with non-vanishing time derivatives. The first and second derivatives give the speed and acceleration of points during the evolution of the wave function. The speed of compression $t<0$ or expansion $t>0$ depends nonlinearly on time and linearly on distance
\begin{equation}
\label{13}
V(r,t)=rt/(\tau_c^2+t^2).
\end{equation}
This expression is a generalized Hubble's law and is valid for any values of time. In the limit of a large enough positive time it leads to the well known Hubble's  law: $V(r,t)\Rightarrow r/t$, $t\gg\tau_c$. In a general case (\ref{13}), the linear dependence of the speed on distance leads to an expression: $V(r,t)/r=t/\left(\tau_c^2+t^2\right)=H(t)$. Physically, function $H(t)$ represents  the speed of expansion of the unit distance in the wave function with  the elapsed time (see FIG.2(a)).
\begin{figure}
\centerline{\includegraphics[width=1.0\columnwidth]{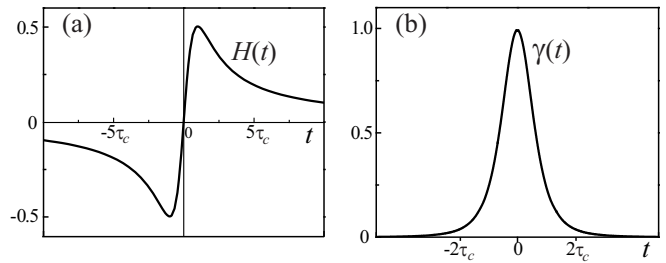}}
\caption{\label{fig2} Graphs of speed $H(t)$ (a), and acceleration  $\gamma(t)$ (b) expansion of unit distance in wave function on the scale of a constant $\tau_c$}
\end{figure}
One more important consequence of the solution (\ref{9}) is the presence of an acceleration $a(r,t)$ of  the evolution of the wave function:
\begin{equation}
\label{15}
a(r,t)=\frac{dV(r,t)}{dt}=r\tau_c^2/\left(\tau_c^2+t^2\right)^2.
\end{equation}
Note that at any moment of time and at any distance the acceleration is a positive quantity. The fact that the speed of expansion depends on the sign of time, whereas the acceleration does not (see FIG.2), is a direct consequence of how the time asymmetry affects the evolution of the Universe. The acceleration (\ref{15}) as well as the speed (\ref{13}), remains  linear functions of distance at any time and increases with the distance to the remote object. Current astronomical observations confirm the presence of an acceleration ~\cite{Riess,Mar,Fre}. However, the existing cosmological models do not provide a reasonable explanation ~\cite{Mar} to this fact without invoking the presence of a hypothetical dark energy ~\cite{Fre} or antigravitation ~\cite{Pee}. According to the expression (\ref{15}), acceleration is a natural consequence of the evolution of spatial geometry of the wave function (\ref{9}).

By comparing expressions (\ref{13}) and (\ref{15}) one can see that in order to determine the value of acceleration of the unit distance (FIG.2.(b))
\begin{equation}
\label{16}
\gamma(t)=a(r,t)/r=\tau_c^2/\left(\tau_c^2+t^2\right)^2
\end{equation}
it is sufficient to know only the speed of expansion $H(t)$ at a given moment of time as $\gamma(t)=H(t)(1-tH(t))/t$.
From this expression it follows that by knowing the exact value of Hubble's constant $(V(r,t_p)/r)=H_0$ at the current moment $t_p$ it is possible to determine the acceleration. The current speed of expansion is measured reliably enough at $H_0\approx 71.0\pm 2.5(km/s)/Mpc$ ~\cite{Seven}. The age of the Universe  from  the same data is approximately $t_p\approx13.75\pm0.13 Gyr$ ~\cite{Seven}. Note that the age of the Universe in (\ref{9}) should be understood as the time which has elapsed since the beginning of expansion. Unfortunately, the wide uncertainty in measurements of these values does not allow one to estimate acceleration $\gamma(t_p)$. It is possible to deduce only that the causality principle in the form of (\ref{11}) imposes restriction on the spread of values in the product: $t_pH_0<1$. Another possibility to determine the acceleration (\ref{16}) is to determine the fundamental constants included in the expression of the wave function (\ref{9}). These constants are the mass $m$ and the initial compression $\rho_0$ (or $\tau_c$ (\ref{12})). In any case, the knowledge of their values is necessary for the complete description of the evolution of the Universe. Importantly, only the product of these constants in the form of a combined constant $\tau_c$ is included in the expressions for the  speed (\ref{13}) and acceleration (\ref{15}). But expression (\ref{11}) imposes a link only between the initial spatial conditions and scale changes of measured values in time. This implies that the system has an additional hidden constant of space-time coupling. This coupling can be determined form the transition from relativistically invariant Klein-Gordon equation $\left(\nabla^{2}-c^{-2}\partial^2_t-m^2c^2\hbar^{-2}\right){\mit\tilde\Psi}=0$ to Schr\"odinger Eq.(\ref{3}) for wave function (\ref{9}) with  broken symmetry of time. By separating the time-dependent factor in the wave function $\mit\tilde\Psi=\mit\Psi e^{-i\omega t}$ and differentiating it %in the Eq.(\ref{17})
with respect to time one obtains
\begin{equation}
\label{18}
\left(\nabla^{2}+\left\{\frac{\omega^2}{c^2}-\frac{m^2c^2}{\hbar^2}\right\}+\frac{2i\omega}{c^2}\frac{\partial}{\partial t}-\frac{1}{c^2}\frac{\partial^2}{\partial t^2}\right){\mit\Psi}=0.
\end{equation}
The expression in curly brackets is identically equal to zero due to $\hbar^2\omega^2=m^2c^4$, and hence Eq.(\ref{18}) can be reduced to the Schr\"odinger Eq.(\ref{3}) if the following condition is satisfied: $\left|c^{-2}\partial_t\partial_t\mit\Psi\right|\ll\left|2m\hbar^{-1}\partial_t\mit\Psi\right|$.
 Thus, the direction in time orientation implies that the amplitude of wave function $\mit\Psi$ changes faster with respect to spatial coordinates than time. This condition cannot be satisfied for the whole coordinate space where unlimited wave function with finite energy exists. Similarly, the condition $c^{-2}\partial_t\partial_t\mit\Psi=2im\hbar^{-1}\partial_t\mit\Psi$  cannot be fulfilled in the whole space as the equation $\nabla^{2}\mit\Psi=0$ has no solutions at which the integrated intensity of a wave $\iiint{\mit\Psi\mit\Psi^*}\,d V$ remains finite within any volume of space. The integrated intensity conservation law (\ref{10}) is universal for the whole space and is not affected by relativistic corrections to the Schr\"odinger equation.
 \begin{figure}
\centerline{\includegraphics[width=0.7\columnwidth]{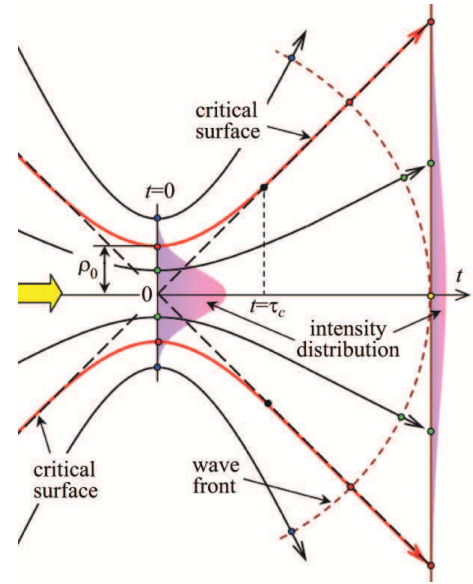}}
\caption{\label{fig3} Pictorial presentation of the fundamental evolution function $\mit\Psi_0(r,t)$. Colour circles show causally dependent points in the space of the function.}
\end{figure}
 Following the above arguments, the Schr\"odinger equation is valid inside a space-time domain whose surface boundary is defined by the condition of equality of spatial and time coordinates in the wave function. In ``empty'' space such a critical surface is just the light cone $r_c(t)=ct$. For the wave function (\ref{9}), there exists only one causally connected (\ref{11}) surface which asymptotically approaches the light cone for $t\gg\tau_c$. It is the surface of a hyperboloid of rotation $r_c(t)=r_c(0)\sqrt{1+t^2/\tau_c^2}$ with the minimum critical radius of $r_c(0)=c\tau_c$. From (\ref{12}) it also evident that the constant of compression and the minimum radius coincide with $r_c(0)=\rho_0$. %Thus, the latent constant of coupling between spatial and time constants in the solution (\ref{9}) is constant of light speed $c$.
  And the universal relationship between constants $\tau_c, \rho_0, m$ in the wave function (\ref{6}) becomes
  \begin{equation}
\label{20}
c\tau_c=\rho_0=2\hbar/(mc).
\end{equation}
 The equation of the critical surface can now be written as $r_c(t)=\rho_0\sqrt{1+t^2/\tau_c^2}=c\sqrt{\tau_c^2+t^2}$. The square of the absolute value of the wave function (\ref{9}) at any point outside the critical surface is $e^2$ times smaller than its maximum value (see FIG.3). Therefore, the solution (\ref{9}) is approximately correct in the whole space except for the critical surface. The fraction of the integrated intensity confined within the critical surface is
$\left.\int_0^{r_c(t)}4\pi r^2{\mit\Psi_0\mit\Psi_0^*}\,d r\right/\int_0^{\infty}4\pi r^2\mit\Psi_0\mit\Psi_0^*\,d r\cong$0.739.
This value  for the wave function $\mit\Psi_0$ (\ref{9}) is an absolute constant at any moment of time.  As a matter of fact, it also defines the portion of matter inside the critical surface. Therefore, it is impossible to identify the radius of the Universe with the radius of the critical surface. The wave function of evolution (\ref{9}) has no real spatial boundaries anywhere on the time axis. When $t\gg\tau_c$ the radius of the critical surface no longer depends on  $\tau_c$. In this case $r_c(t)\cong ct$ and, provided that the age of the Universe is estimated correctly ~\cite{Seven}, the critical radius of the current Universe is $r_c(t_p)\approx1.3\times10^{26}m$. Speed of expansion of points (\ref{13}) on the critical surface asymptotically approaches the speed of light $V_c(t)=\rho_0t/\left(\tau_c\sqrt{\tau_c^2+t^2}\right)\Rightarrow c$ (see FIG.4). Physically, it means that causally connected space of events inside the critical surface cannot expand faster than the speed of light.
\begin{figure}
\centerline{\includegraphics[width=0.6\columnwidth]{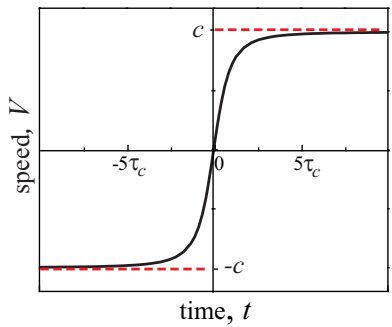}}
\caption{\label{fig4} Time dependence of expansion speed $V(t)$ on the critical radius $r_c$.}
\end{figure}
Interestingly, in the large time limit it is not necessary to know the values of constants $\tau_c, \rho_0, m$ to define the size and speed of expansion of the wave function (\ref{9}). The relationship between the constants (\ref{20}) allows their variations provided that the condition $m\rho_0=2\hbar/c$ or $m\tau_c=2\hbar/c^2$ remains valid. Such variations do not influence the speed of expansion and the size of the Universe in the present time when $t_p\gg\tau_c$. However, it is imperative to know the value of at least one of the constants to determine the acceleration. Indeed, according to Eq.(\ref{16}) the current acceleration of expansion is given by $\gamma(t_p)\cong\tau_c^2/t_p^4=4\hbar^2/(m^2c^4t_p^4)$  and in order to find its value it would be sufficient to determine the mass of the Universe. Some care is required however in defining what one  means by the total mass of the Universe. If it mass has a value of $m\sim0.8 - 1.5\times 10^{53}kg$, which is derived from  the density of the observable substance ~\cite{Teor,Mo}, then the current value of acceleration of expansion is vanishingly small: $\gamma(t_p)\sim10^{-254}(m/s^2)/Mpc$. The acceleration (\ref{15}) on the critical surface has dropped by $a(r_c,0)/a(r_c,t_p)\simeq10^{364}$ during time $t_p$. The above value for the total mass also implies that the compression constant is $\rho_0\sim10^{-96}m$ . This size is many orders below the critical Planck length and seems to be  nonphysical. On the other hand, if one assumes the critical radius of the Universe to be equal to the Planck length $\rho_0=l_p$ at the moment $t=0$, then, according to expression (\ref{20}), constant $\tau_c=\tau_p$ is the Planck time, but the mass of the Universe becomes only two Planck masses $m=2m_p$. At a first glance this result seems improbable. But, this mass may not be connected to the mass of the observable substance. The meaning of the mass of the Universe remains obscure. Nevertheless, the probability Planck's values of the Universe mass does not contradict to the modern physical understanding. The total mass of the Universe may consists of positive and negative masses of the matter ~\cite{Bond}, but the conservation laws remain unbroken. QM does not forbid existence of a locally mass-negative region of space-time too ~\cite{Mor}. Moreover,the concept of the negative mass of  of the physical vacuum implicitly contains in the equations of P. Dirac and P. Higgs ~\cite{Dir,Higgs}. In this case it is quite admissible that the balance of mass of the Universe is tipped to a positive side by only double Planck mass. If this hypothesis is correct, the current acceleration has a value  $\gamma(t_p)\approx2.5\times10^{-135}(m/s^2)/Mpc$. The acceleration (\ref{15}) on the critical surface has dropped over the time $t_p$ by factor $a(r_c,0)/a(r_c,t_p)\simeq5\times10^{182}$. Even though the estimated present accelerations are vanishingly small, the difference in their orders remains huge. An experimental estimate of the acceleration of expansion of the Universe at any stage of its evolution, even with an error of several orders, would provide an unambiguous answer regarding the validity of the proposed ere mass of the Universe.

In this work possible modulating functions defining the local geometry of the surface of the wave function are not shown explicitly. Similarly to the solutions given by (\ref{5}), the general solution of Eq.(\ref{3}) can be presented in the form of a product of the background (\ref{9}) and modulating functions $\mit\Psi=F(r,t)\mit\Psi_0$. The functions $F(r,t)$ should satisfy 3D space Helmholz wave equation or 3D Laplace's source equation: $\nabla^{2}F(r,t)=0$. They determine the local topology of space inside the basic function (\ref{9}). The solutions of 3D Laplace equation are well known polynomial spherical functions. In QM they describe discrete properties of the particles ~\cite{Fermi}. Note additionally that in special cases the modulating functions can assume a vector representation only, by analogy with EMW.

In conclusion we would like to emphasize that the exponential nature of the evolution of the Universe is not new and fits the Inflation model ~\cite{Guth, Lin, Barr}. Unlike the Inflation model, the proposed QM model for the evolution of the Universe does not have the initial singularity, operates with finitely integrable functions and for the first time explains the direction of time arrow. These issues remain unresolved in the Inflation model ~\cite{Page}. At the same time, there are a number of questions pertaining to our model which have to be answered. For instance, the exact physical meaning of the very evolution function (\ref{9}) is not completely clear. Most likely its squared absolute value describes a probability density of the free matter distribution in space at each moment of time. One may argue that this interpretation disagrees with the observed homogeneous distribution of a substance in the Universe. The function (\ref{9}), however, may be completely irrelevant to the distribution of the observable substance as the latter is described by the external source functions $F(r,t)$ rather than by the evolution function (\ref{9}). For example, it was mentioned earlier that the singular source functions (\ref{7}) in (2+1)D Schr\"odinger Eq.(\ref{2}) are not related to the intensity distribution of the background wave function (\ref{4}) and can fill total uniformly the whole space of the wave function (\ref{4}). This can be clearly seen on the example of speckle fields (FIG. 1), where the filling of the space with the sources of singularities is governed by the carrier wavelength and not by the distribution function of its global intensity ~\cite{Den}.

The most remarkable aspect of the proposed model is that gravitation has nothing to do with the evolution of the Universe, because neither the equations for gravitational field nor the gravitational constant are explicitly included into the fundamental solution. In other words, the space of events, which is described by the fundamental evolution function (\ref{9}), becomes the prerequisite for the existence of gravitation. This follows from the fact that substance sources which are described by the modulating functions $F(r,t)$ are external with respect to the fundamental function $\mit\Psi_0$. They are responsible for the local topology of the space confined inside the space of the fundamental function (9) and do not influence the general character of evolution. Physically, the background wave function (\ref{9}), which is the fundamental solution of the time-dependent Schr\"odinger equation (\ref{3}), defines a common space where all objects described by the modulating functions can exist.

I thank Dr. C. Hnatovsky and Prof. W. Krolikowski for useful discussions.

\end{document}